\documentclass[10pt]{article}
\usepackage[reqno]{amsmath}
\usepackage{amssymb,amsfonts,amsthm,upref,graphics,graphicx}

\newcommand{\R}{{\mathbb R}}

\newcommand{\N}{{\mathbb N}}
\newcommand{\Oo}{{\cal O}}

\newtheorem{theo}{{\bf Theorem}}[section]
\newtheorem{defn}[theo]{{\bf Definition}}
\newtheorem{cor}[theo]{{\bf Corollary}}

\newtheorem{prop}[theo]{{\bf Proposition}}

\numberwithin{equation}{section}

\author{Alain Bourget\thanks{\textbf{Mailing address}: Department of Mathematics, California State
University (Fullerton), McCarthy Hall 154, Fullerton CA 92834 (US).} \\ Department of Mathematics \\ California
State University, Fullerton }

\title{Spectral Density of Jacobi Matrices with small deviations} 

\date{}

\begin{document}

\maketitle

\begin{abstract}
We present several new asymptotic trace formulas for Jacobi matrices whose coefficients satisfy a small deviation condition. Our results extend most of the existing trace formulas for Jacobi matrices.
\end{abstract}

\medskip


\section{Introduction}
Let $\mathbf{a}^k=(a_1^k,...,a_k^k) \in \R^k$ and $\mathbf{b}^k=(b_1^k,...,b_k^k) \in \R^k$. By a
Jacobi matrix, we mean a real, symmetric, tridiagonal matrix of the form
\begin{equation}\label{finitejacobi}
J(\mathbf{a}^k,\mathbf{b}^k) = \left(
\begin{array}{ccccc}
a_1^k & b_1^k & 0 & \cdots & 0 \\
b_1^k & a_2^k & b_2^k & \ddots & \vdots \\
0 & b_2^k & a_3^k & \ddots & 0 \\
\vdots & \ddots & \ddots & \ddots & b_{k-1}^k \\
0 & \cdots & 0 & b_{k-1}^k & a_k^k \\
\end{array}
\right).
\end{equation}

Jacobi matrices have a wide range of applications in mathematical sciences. In physics for instance, they
naturally appear in the study of random matrices and discrete Schr\"odinger operators \cite{MR883643}. In
statistics, they provide a useful tool to study stochastic processes such as the birth-death process and random
walks. In classical analysis, they play an important role in the study of orthogonal polynomials \cite{deift96, MR903848}.

The spectral properties of $J(\mathbf{a}^k,\mathbf{b}^k)$ are well-known and can be found in many texts; a good reference is e.g. \cite{deift96}. If $M$ denotes the quantity
\[ M=\sqrt{3} \left[ \max \limits_{1 \leq i \leq k} |a_i^k| +  \max \limits_{1 \leq i \leq k} |b_i^k| \right], \]
then one can easily verify that $\sigma(J(\mathbf{a}^k,\mathbf{b}^k))$, the spectrum of $J(\mathbf{a}^k,\mathbf{b}^k)$, consists of $k$ real simple eigenvalues lying inside the interval $[-M,M]$.

Except for few special cases, it is in general impossible to obtain explicit expressions for the eigenvalues of an arbitrary Jacobi matrix. However, an important special case for which the spectrum is explicitly known is the case of   tridiagonal Toeplitz matrices.  These matrices have the form
\begin{equation}
T_k(a,b) = \left(
\begin{array}{ccccc}
a & b & 0 & \cdots & 0 \\
b & a & b & \ddots & \vdots \\
0 & b & a & \ddots & 0 \\
\vdots & \ddots & \ddots & \ddots & b \\
0 & \cdots & 0 & b & a \\
\end{array}
\right)
\end{equation}
for some $a \in \R$ and $b>0$. The eigenvalues of $T_k(a,b)$ are well-known and are given by
\[ \lambda_j^k = a +2b \cos \left( \frac{j}{k}\pi \right) \qquad \text{ for } j=1,...,k. \]
Using Riemann sums, it is then straightforward to derive the following asymptotic trace formula
\begin{eqnarray*}
\lim_{k \to \infty} \frac{1}{k} \text{Trace}[\phi(T_k(a,b))] & = &
\lim_{k \to \infty} \frac{1}{k} \sum_{j=1}^k \phi \left( a+2b\cos \left( \frac{j}{k} \pi \right) \right) \\
& = & \frac{1}{\pi} \int_0^{\pi} \phi(a+ 2b \cos x) \, dx
\end{eqnarray*}
for any continuous function $\phi$ on the interval $[a-2b,a+2b]$. This trace formula was successfully used to obtain new ones for Jacobi matrices that are small perturbation of a Toeplitz matrix. For instance, this approach can be used to derive Nevai's result \cite{ne79} on the density of the zeros of orthogonal polynomials that belongs to the $M(a,b)$-class.

In this paper, we continue our investigation initiated in our recent work \cite{agbo09}. We start by deriving two asymptotic trace formulas for the moments that extend the ones given in Lemma 2.3 of \cite{agbo09}. In the third section, we derive our main results. Namely, if we assume the bounded sequences $(\mathbf{a}^k)_{k}, (\mathbf{b}^k)_{k}$ to satisfy the small-deviation conditions of Definition \ref{def1} and  if $(\mathbf{a}^k,\mathbf{b}^k)_{k}$ is $\mu$-distributed as in the sense of Definition \ref{mu-distributed}, then we have
\begin{equation} \label{main}
\lim_{k \to \infty} \frac{1}{k} \text{Trace}[\phi(J(\mathbf{a}^k,\mathbf{b}^k))] = \frac{1}{\pi} \int_{[0,1]^2} \int_0^\pi \phi(x+2y\cos t) \, dt \, d\mu(x,y)
\end{equation}
for any $\phi \in C[-3,3]$. We also present similar results for unbounded sequences. In addition, we show in the fourth section of the paper that for any given sequence $(\mathbf{a}^k)_k, \ (\mathbf{b}^k)_{k}$ satisfying the small-deviation conditions of Definitions \ref{def1} and  \ref{unbounded}, one can always find a subsequence and probability measure $\mu$ for which a trace formula similar to \eqref{main} holds.

\section{Moments}

We start by proving a trace formula for the moments of $J(\mathbf{a}^k,\mathbf{b}^k)$ when the sequences $(\mathbf{a}^k)_k$ and $(\mathbf{b}^k)_k$ are bounded and satisfy the small deviation condition below.  We also present results when the sequences are unbounded.

The first thing consists of defining in precise terms what we mean for a sequence $(\mathbf{a}^k)_k$ with  $\mathbf{a}^k=(a_1^k,...,a_k^k) \in \R^k$ to satisfy the required small deviation condition.

\begin{defn} \label{def1}
We say that the sequence $(\mathbf{a}^k)_k$ belongs to $\mathcal{S}$ if it satisfies the following two conditions:
\begin{itemize}
\item[(i)] $\sum_{i=1}^{k-1} |a_{i+1}^k-a_{i}^k|= o(k)$ as $k \to \infty$,
\item[(ii)] $0\leq a_i^k  \leq 1$ for all $i=1,...,k$.
\end{itemize}
\end{defn}

Note that (ii) holds without loss generality for any bounded sequence. Indeed, if $|a_i^k|<M$ for all $i \leq k$ and $k \in \N$, then the normalized sequence $(\mathbf{\tilde{a}}^k)_k$ with $\tilde{a}_i^k=1/2+a_i^k/2M$ satisfies $(ii)$.

\medskip

\noindent \textbf{Example 1:} Let $(a_k)_k$ be any convergent sequence in $[0,1]$. If we let $a_i^k=a_i$ for $i \leq k$, then it is easy to see that $(\mathbf{a}^k)_k \in \mathcal{S}$.

\medskip

\noindent \textbf{Example 2:} As an example of a sequence $(\mathbf{a}^k)_k \in \mathcal{S}$ that is not necessarily obtained from a convergent sequence, consider a bounded sequence $(\mathbf{a}^k)_k$ that satisfies for any $0<\delta<1$:
$$ \frac{\#\{ 1 \leq j \leq k: |a_{j+1}^k-a_j^k| = \Oo(k^{-\delta}) \}}{k} \to 1 \qquad \text{ as } k \to \infty.$$
One can easily verify that $(\mathbf{a}^k)_k$ satisfies condition (i) of Definition \ref{def1}. For instance, any uniformly distributed sequence $(\mathbf{a}^k)_k$ modulo one with discrepancy
$$D_k =\max_{1 \leq i \leq k} \left\{ \left| a_i^k-\frac{i}{k} \right|, \left|a_i^k - \frac{i-1}{k} \right| \right\} = \Oo(k^{-\delta})$$
falls into this category \cite{MR0419394}.

\medskip

Our first trace formula is concerned with the moments of Jacobi matrices $J(\mathbf{a}^k,\mathbf{b}^k)$ whose defining sequences $(\mathbf{a}^k)_k$ and $(\mathbf{b}^k)_k$ belong to $\mathcal{S}$.

\begin{prop} \label{prop 1}
Let $(\mathbf{a}^k)_k, (\mathbf{b}^k)_k$ be two sequences in $\mathcal{S}$. For any $n \in \N$, we have
\begin{equation} \label{trace000}
\text{Trace}\left[ J^n(\mathbf{a}^k,\mathbf{b}^k) \right]=\sum_{j=0}^{\lfloor n/2 \rfloor} \frac{n!}{(j!)^2(n-2j)!} \sum_{i=1}^k (a_i^k)^{n-2j} (b_i^k)^{2j} +  o(k).
\end{equation}
\end{prop}

\noindent \textit{Proof:} We write $J_k:=J(\mathbf{a}^k,\mathbf{b}^k)$ as the sum of three matrices $$J_k=L_k+D_k+L^T_k$$ where $D_k=\text{diag}(a_1^k,...,a_k^k)$ and $L_k$ is the lower triangular matrix given by
\begin{equation}
L_k = \left(
\begin{array}{ccccc}
0 & 0 & 0 & \cdots & 0 \\
b_1^k & 0 & 0 & \ddots & \vdots \\
0 & b_2^k & 0& \ddots & 0 \\
\vdots & \ddots & \ddots & \ddots & 0 \\
0 & \cdots & 0 & b_{k-1}^k & 0 \\
\end{array}
\right).
\end{equation}

For any $n \in \N$, $J^n_k=(L_k^T+D_k+L_k)^n$ is the sum of $3^n$ matrix monomials of the form $$A_1A_2 \cdots A_n$$ with $A_j$ being either $L_k$, $D_k$ or $L_k^T$. One can easily verify that the monomials having a different number of $L_k$ and $L_k^T$ in their expressions  have zero trace. Therefore, it suffices to consider those having the same number of $L_k$ and $L_k^T$.

The assumptions on the sequences $(\mathbf{a}^k)_k$ and $(\mathbf{b}^k)_k$ imply that when we permute any two consecutive matrices in the product $A_1 A_2 \cdots A_n$, the trace of the resulting expression differ from the trace of the original one by $o(k)$.  In other words, we have
\[ \frac{1}{k} \text{Trace}[A_1 \cdots A_i A_{i+1} \cdots A_n] = \frac{1}{k} \text{Trace}[A_1 \cdots A_{i+1} A_i \cdots A_n ] + o(1). \]
To see this, let us consider the case where $A_i=L^T_k$ and $A_{i+1}=L_k$; the other cases can be handled in a similar manner. We have that
\begin{eqnarray*}
  A_1A_2 \cdots A_n & = & A_1 A_2 \cdots A_{i+1}A_i \cdots A_n \\
    & & + A_1 A_2 \cdots A_{i-1} [A_i,A_{i+1}] A_{i+2} \cdots A_n.
\end{eqnarray*}
By an elementary property of the trace, we also have
\begin{multline*}
  \text{Trace}[ A_1 A_2 \cdots A_{i-1} [A_i,A_{i+1}] A_{i+2} \cdots A_n] \\= \text{Trace}[ [A_i,A_{i+1}]A_{i+2}\cdots A_n A_1 \cdots A_{i-1} ]
\end{multline*}
The product $A_{i+2} \cdots A_n A_1 \cdots A_{i-1}$ is a matrix whose diagonal elements are bounded, while the commutator matrix $[A_i,A_{i+1}]$ is a diagonal matrix whose elements are $(b_1^k)^2,(b_2^k)^2-(b_1^k)^2,...,(b^k_k)^2-(b_{k-1}^k)^2,(b^k_k)^2$. Under the assumptions that $(\mathbf{a}^k)_k$ and $(\mathbf{b}^k)_k$ belong to $\mathcal{S}$, we easily deduce
\begin{eqnarray*}
 \text{Trace}[ [A_i,A_{i+1}] A_{i+2} \cdots A_n A_1 A_2 \cdots A_{i-1} ]  & \leq & |b_1^k|^2 + \sum_{j=1}^{k-1} |b_{j+1}^k-b_j^k|^2 + |b_k^k|^2 \\
 & = & o(k).
\end{eqnarray*}
Consequently, it suffices to consider terms of the form $(L_kL^T_k)^{2j} D_k^{n-2j}$ for which the trace can easily be computed. Indeed, we have
\[  \text{Trace}[(L_kL^T_k)^{2j} D_k^{n-2j}]=\sum_{i=1}^k (a_i^k)^{n-2j} (b_i^k)^{2j}. \]
Since there are $\frac{n!}{(j!)^2(n-2j)!}$ monomials containing $j$ matrices $L_k$ and $L_k^T$ and $n-2j$ matrices $D_k$, we finally obtain
\begin{eqnarray*}
\lefteqn{\frac{1}{k} \text{Trace}\left[ J^n(\mathbf{a}^k,\mathbf{b}^k) \right]}\\
& = & \frac{1}{k}\sum_{j=0}^{\lfloor n/2 \rfloor} \frac{n!}{(j!)^2(n-2j)!} \text{Trace}[(L_kL^T_k)^{2j} D_k^{n-2j}] + o(1) \\
& = & \frac{1}{k}\sum_{j=0}^{\lfloor n/2 \rfloor} \frac{n!}{(j!)^2(n-2j)!} \sum_{i=1}^k (a_i^k)^{n-2j} (b_i^k)^{2j} + o(1)
\end{eqnarray*}
as desired. $\qed$

\bigskip

The boundedness condition $(ii)$ in Definition \ref{def1} can be removed if we slightly strengthen the first condition. More precisely, we have:

\begin{defn} \label{unbounded}
We say that the sequence $(\mathbf{a}^k)_k$  belongs to $\mathcal{S'}$ if it satisfies the following two conditions: For any $\delta \in (0,1)$,
\begin{itemize}
\item[(i')] $\sum_{i=1}^{k-1} |a_{i+1}^k-a_{i}^k|= \Oo(k^{1-\delta})$ as $k \to \infty$,
\item[(ii')] $\max \limits_{1 \leq i \leq k} |a_i^k|= \Oo(\log k)$ as $k \to \infty$
\end{itemize}
\end{defn}

It is readily seen that a similar proof as the one of Proposition \ref{prop 1} holds when the sequences $(\mathbf{a}^k)_k $ and $(\mathbf{b}^k)_k$ are both in $\mathcal{S}'$. We state this result as our second trace formula.

\begin{prop} \label{prop 2}
Let $(\mathbf{a}^k)_k, \,(\mathbf{b}^k)_k \in \mathcal{S}'$. For any $n \in \N$, we have
\begin{equation} \label{trace00}
\text{Trace}\left[ J^n(\mathbf{a}^k),(\mathbf{b}^k) \right]= \sum_{j=0}^{\lfloor n/2 \rfloor} \frac{n!}{(j!)^2(n-2j)!} \sum_{i=1}^k (a_i^k)^{n-2j} (b_i^k)^{2j} +  o(k).
\end{equation}
\end{prop}

\section{Main trace formulas: The $\mu$-distributed case}

We now derive our main asymptotic trace formulas for Jacobi matrices whose defining sequences $(\mathbf{a}^k)_k$ and $(\mathbf{b}^k)_k$ are distributed according to some probability measure. We consider different cases depending on if the sequences $(\mathbf{a}^k)_k$ and $(\mathbf{b}_k)^k$ are bounded, unbounded or monotone.

\subsection{The bounded case}

Our first results are concerned with compactly supported probability measure $\mu$ on $\R^2$. After normalization, we may assume without loss of generality that the support is contained in $I^2:=[0,1] \times [0,1]$.

\begin{defn}  \label{mu-distributed}
We say that a sequence $(\mathbf{a}^k,\mathbf{b}^k)_k$ with $a_j^k$ and $b_j^k$ in $[0,1]$ for all $j=1,...,k$ is $\mu$-distributed if for any continuous function $\psi$ on $[0,1]^2$, one has
\[ \lim_{k \to \infty} \frac{1}{k} \sum_{j=1}^k \psi(a^k_j,b^k_j) = \int_{I^2} \psi(x,y) \ d\mu(x,y). \]
\end{defn}

\noindent \textbf{Example 3:} Consider the sequences $(\mathbf{a}^k)_k$ and $(\mathbf{b}^k)_k$ that are convergent, i.e. $\mathbf{a}^k=(a_1,...,a_k)$ and $\mathbf{b}^k=(b_1,...,b_k)$ with $a_k \to a$ and $b_k \to b$. It is easy to see that $(\mathbf{a}^k,\mathbf{b}^k)_k$ is $\mu$-distributed with $\mu=\delta_a \times \delta_b$, the product of the Dirac measures at $a$ and at $b$.

\medskip

\noindent \textbf{Example 4:} Let $a,b:[0,1] \to [0,1]$ be two continuous functions. We define the sequences $\mathbf{a}^k$ and $\mathbf{b}^k $ by
$$\mathbf{a}^k=(a(1/k),a(2/k),...,a((k-1)/k),a(1))$$
and
$$\mathbf{b}^k=(b(1/k),b(2/k),...,b((k-1)/k),b(1)).$$
It follows that $(\mathbf{a}^k,\mathbf{b}^k)_k$  is $m_f$-distributed  where $m_f$ is the probability distribution associated to the random variable $f(x)=(a(x),b(x))$, i.e.
\[ \int_{[0,1]^2} \psi(x,y) \, dm_{f} (x,y) = \int_0^1 \psi(a(x),b(x)) \, dx \]
for any $\psi \in C(I^2)$.

\medskip

In our first theorem of this section, we give a trace formula for Jacobi matrices with bounded sequences that are $\mu$-distributed.

\begin{theo} \label{theo 1}
Let $\mu$ be a probability measure on $I^2$. Suppose that $(\mathbf{a}^k, \mathbf{b}^k)_k$ is a $\mu$-distributed sequence with $(\mathbf{a}^k)_k, \ (\mathbf{b}^k)_k \in \mathcal{S}$. Then, we have
\begin{equation} \label{trace1}
\lim_{k \to \infty} \frac{1}{k} \text{Trace}[\phi(J(\mathbf{a}^k,\mathbf{b}^k))] = \frac{1}{\pi} \int_{I^2} \int_0^\pi \phi(x+2y\cos t) \, dt \, d\mu(x,y)
\end{equation}
for any $\phi \in C[-3,3]$.
\end{theo}

\noindent \textit{Proof:} For any nonnegative integer $n$, Proposition \ref{prop 1} implies
\begin{multline} \label{trace3}
\frac{1}{k} \text{Trace}\left[ J^n(\mathbf{a}^k,\mathbf{b}^k) \right] \\
= \sum_{j=0}^{\lfloor n/2 \rfloor} \frac{n!}{(j!)^2(n-2j)!} \int_{I^2} x^{n-2j}y^{2j} \, d\mu(x,y) + o(1).
\end{multline}

In order to replace the sum $\sum_{j=0}^{\lfloor n/2 \rfloor}$ by the sum over all multi-indices $\alpha=(\alpha_1,\alpha_2,\alpha_3)$ with $|\alpha|=n$, we introduce the function $sinc:\R \to \R$  defined by
\begin{equation*}
\text{sinc}( \xi) = \frac{1}{2} \int_{-1}^1 e^{i\pi t \xi} \, dt.
\end{equation*}
Note that $\text{sinc}(0)=1$ while $\text{sinc}(\pi x)=0$ when $x$ is a non-zero integer. Using this function, \eqref{trace3} becomes
\begin{multline*}
\frac{1}{k} \text{Trace}\left[ J^n(\mathbf{a}^k,\mathbf{b}^k) \right] \\
= \frac{1}{2}  \int_{I^2} \int_{-1}^1 \sum_{|\alpha|=n} \binom{n}{\alpha} e^{i\pi t(\alpha_1-\alpha_2)} x^{\alpha_3}y^{\alpha_1+\alpha_2} \, d\mu(x,y) \, dt + o(1).
\end{multline*}
The multinomial theorem and a simple change of variables then yield
\begin{equation} \label{trace5}
\frac{1}{k} \text{Trace}\left[ J^n(\mathbf{a}^k,\mathbf{b}^k) \right]
= \frac{1}{\pi} \int_{I^2} \int_0^\pi (x+2y\cos(t))^n \, dt \, d\mu(x,y)  + o(1).
\end{equation}

This establishes the result for the moments of $J(\mathbf{a}^k,\mathbf{b}^k)$. By linearity, the result also holds for polynomials of arbitrary degree. Now let $\phi$ be a continuous function on $[-3,3]$. Note that
$$ \{ x+2y\cos(t): \, (x,y) \in I^2, \, t \in [0,\pi] \} \subseteq [-1,3]$$
and $\sigma(J(\mathbf{a}^k,\mathbf{b}^k)) \subseteq [-2\sqrt{3},2\sqrt{3}]$.
By Weierstrass Approximation Theorem, there is a polynomial $P$ such that $\|\phi-P\|_{\infty} < \epsilon/3$ on $[-3,3]$. In particular, it implies that
\begin{equation} \label{trace7}
\left| \frac{1}{k} \text{Trace}[\phi(J(\mathbf{a}^k,\mathbf{b}^k))] -  \frac{1}{k} \text{Trace}[P(J(\mathbf{a}^k,\mathbf{b}^k))] \right| \leq \epsilon/3,
\end{equation}
together with
\begin{multline} \label{trace8}
\left| \frac{1}{\pi} \int_{I^2} \int_0^\pi \phi(x+2y\cos(t)) \, dt \, d\mu(x,y) \right. \\
\left. -\frac{1}{\pi} \int_{I^2} \int_0^\pi P(x+2y\cos(t)) \, dt \, d\mu(x,y) \right| \leq \epsilon/3.
\end{multline}

Finally, \eqref{trace5} shows that we can choose $k$ large enough so that
\begin{equation} \label{trace9}
\left| \frac{1}{k} \text{Trace}\left[ P(J(\mathbf{a}^k,\mathbf{b}^k)) \right]
-\frac{1}{\pi} \int_{I^2} \int_0^\pi P(x+2y\cos(t)) \, dt \, d\mu(x,y) \right| \leq \epsilon/3.
\end{equation}
The conclusion of the theorem follows by combining \eqref{trace7}, \eqref{trace8}, and \eqref{trace9}. $\qed$

\medskip

\noindent \textbf{Example 5:} Let $(\mathbf{a}^k)_k$ and $(\mathbf{b}^k)_k$ be convergent sequences as in Example 3. From Theorem \ref{theo 1}, we easily deduce that
\begin{eqnarray*}
\lim_{k \to \infty} \frac{1}{k} \, \text{Trace}[\phi(J(\mathbf{a}^k,\mathbf{b}^k))] & = & \frac{1}{\pi} \int_0^\pi \phi(a+2b \cos t) \, dt \\
& = & \frac{1}{\pi} \int_{a-2b}^{a+2b} \phi(t) \frac{dt}{\sqrt{(a+2b-u)(u-a+2b)}}.
\end{eqnarray*}
This type of Jacobi matrices naturally arise for orthogonal polynomials that belong to the $M(a,b)$ class introduced by Nevai \cite{ne79}. One can use this result to easily compute the asymptotic distribution of the zeros of many classical orthogonal polynomials such as the Jacobi ($a=1/4$, $b=1/2$), Chebyshev ($a=0$, $b=1/2$), Legendre $(a=0, b=1/2)$ and Gegenbauer $(a=0, b=1/2)$ polynomials.

\medskip

\noindent \textbf{Example 6:} In computing the asymptotic distribution of the zeros of Van Vleck polynomials \cite{bosh08}, one is led to consider a Jacobi matrix of the form
\begin{equation}\label{finitejacobi}
J_k = \left(
\begin{array}{ccccc}
a_1 &  b_1  & 0 & \cdots & 0 \\
b_1 & a_2  & b_2 & \ddots & \vdots \\
0 & b_2 & a_3 & \ddots & 0 \\
\vdots & \ddots & \ddots & \ddots & b_{k-1} \\
0 & \cdots & 0 & b_{k-1} & a_k \\
\end{array}
\right).
\end{equation}
where the entries are up to some constants given by
\[ a_j=\frac{(j-1)}{k}  \text{ and } b_j=\frac{j\sqrt{1-(j/k)^2}}{k}. \]

It is easy to see that the sequences $(a_k)_k$ and $(b_k)_k$ are both in $\mathcal{S}$, and the sequence $(a_k,b_k)_k$ is $\mu$-distributed with
\[ \mu(x,y) = \frac{\delta_0(y-\sqrt{x(1-x)})}{2\sqrt{x}}. \]
It then follows from Theorem \ref{theo 1} that
\begin{equation*}
\lim_{k \to \infty} \frac{1}{k} \text{Trace}[\phi(J_k)] = \frac{1}{\pi} \int_0^1 \int_0^\pi \phi(x+2\sqrt{x(1-x)} \cos t) \, dt \, \frac{dx}{2 \sqrt{x}}.
\end{equation*}

\medskip

In a recent paper, Kuiljaars and Serra-Cappizano \cite{MR1866252} proved a general result regarding the asymptotic distribution of the eigenvalues of Jacobi matrices. Their results generalize earlier ones obtained by Kuiljaars and Van Assche \cite{kuva99} and Geronimo, Harrell II and Van Assche \cite{MR956176}. We give here a new proof of their result based on our previous trace formula in the generic case $|a_j^k| \leq 1$ and $0< b_j^k \leq 1$.

\begin{cor} (Kuiljaars-Serra Cappizano) Let $(J(\mathbf{a}^k,\mathbf{b}^k))_k$ be a sequence of Jacobi matrices and suppose we can choose two bounded measurable functions $a,b:[0,1] \to [0,1]$ satisfying the conditions: For any $\epsilon>0$,
\begin{equation} \label{ksc cond 1}
|\{ s \in [0,1]: |a_{\lceil sk \rceil}^k - a(s) | \geq \epsilon \}| \to 0 \text{ as } k \to \infty,
\end{equation}
\begin{equation} \label{ksc cond 2}
|\{ s \in [0,1]: |b_{\lceil sk \rceil}^k - b(s) | \geq \epsilon \}| \to 0 \text{ as } k \to \infty.
\end{equation}
Then, we have
\begin{equation*}
\lim_{k \to \infty} \frac{1}{k} \, \text{Trace}[\phi(J(\mathbf{a}^k,\mathbf{b}^k))] = \frac{1}{\pi} \int_0^1 \int_0^\pi \phi(a(x)+2b(x) \cos y) \, dx \, dy
\end{equation*}
for any $\phi \in [-3,3]$.
\end{cor}

\noindent \textit{Proof:} For $k$ large enough, conditions \eqref{ksc cond 1} and \eqref{ksc cond 2} together with  Lusin's Theorem imply that we can choose the functions $a$ and $b$ to be continuous on $[0,1]$ except for a set of measure $1/k$ and for which the additional conditions
\[ \#\{j \leq k :|a_j^k-a(j/k)| \geq \epsilon \} =o(k) \]
\[  \#\{j \leq k :|b_j^k-b(j/k)| \geq \epsilon \} =o(k) \]
hold for any given $\epsilon>0$. It easily follows that for any continuous function $\psi$ on $[0,1]^2$,
\begin{equation} \label{trace15}
\lim_{k \to \infty} \frac{1}{k} \sum_{j=1}^k \psi(a_j^k, b_j^k)  =  \int_0^1 \psi(a(x),b(x)) \, dx.
\end{equation}
If we introduce the map $f:[0,1] \to [0,1]^2$ defined by $f(x)=(a(x),b(x))$, then it follows as in Example 4 that the sequence $(\mathbf{a}^k,\mathbf{b}^k)_k$ is $m_f$-distributed. The conclusion is then an immediate consequence of Theorem \ref{theo 1}. $\qed$

\medskip

In the next result, we derive a simple extension of Example 5. Indeed, we consider sequences $(\mathbf{a}^k)_k$ and $(\mathbf{b}^k)_k$ that have a finite number of accumulation points.

\begin{cor}
Let $(\mathbf{a}^k)_k$ and $(\mathbf{b}^k)_k$ be two sequences in $\mathcal{S}$. Let $\Omega=[\omega_1,...,\omega_n]$ be an $n$ vector for which  $0 \leq \omega_i \leq 1$ and
$$|\Omega|:=\sum_{i=1}^n \omega_i=1.$$
Suppose there exists two $n$-tuple $(\alpha_1,...,\alpha_n)$ and  $(\beta_1,...,\beta_n)$ such that for any $1\leq p \leq n$, and $\epsilon>0$,
\begin{equation} \label{cond a}
\frac{\#\{ i : |a_i^k-a_p|< \epsilon \text{ and } |b_i^k-b_p| < \epsilon \}}{k} \to \omega_p  \ \text{ as } k \to \infty.
\end{equation}
Then, for every continuous function $\phi$ on $[-3,3]$, one has
\begin{equation}
\lim_{k \to \infty} \frac{1}{k} \sum_{j=1}^k \text{Trace}[\phi(J(\mathbf{a}^k,\mathbf{b}^k))] = \frac{1}{\pi} \int_0^\pi \int_{[0,1]^2} \phi(x+2y\cos t) \,d\omega(x,y) \,dt
\end{equation}
where $\omega$ is the probability measure given by the weighted sum of delta measures, i.e.
\begin{equation}
\omega(x,y)=  \sum_{j=1}^n \omega_i \, \delta_{a_i}(x) \times \delta_{b_i}(y).
\end{equation}
\end{cor}

\noindent \textit{Proof:} It is readily seen that both sequences satisfy the assumptions of Theorem \ref{theo 1} with $\mu$ given by $\omega$ as defined in the statement of the theorem. $\qed$

\medskip

As a simple consequence of the previous result, we have the following result also due to Kuiljaars and Serra Capizzano \cite{MR1866252} when the sequences $(\mathbf{a}^k)_k$ and $(\mathbf{b}^k)_k$ have a single accumulation point.

\begin{cor} (Kuiljaars-Serra Capizzano) \label{cor ks}
Let $(a_k)_k$ and $(b_k)_k$, $b_k>0$, be two bounded sequences as in Example 1, i.e. $(\mathbf{a}^k)_k$ is the sequence defined by $\mathbf{a}^k=(a_1,...,a_k)$ and similarly for $(\mathbf{b}^k)_k$. Suppose there exist real constant $a$ and $b>0$ such that for every $\epsilon>0$,
\begin{equation}
\#\{j\leq k: |a_j-a| \geq \epsilon \} = o(k) \ \text{ as } k \to \infty,
\end{equation}
and
\begin{equation}
\#\{j\leq k: |b_j-b| \geq \epsilon \} = o(k) \ \text{ as } k \to \infty.
\end{equation}
Then, for every continuous function $\phi$ on $[a-2b,a+2b]$,
\begin{equation}
\lim_{k \to \infty} \frac{1}{k} \sum_{j=1}^k \text{Trace}[\phi(J(\mathbf{a}^k,\mathbf{b}^k))] = \frac{1}{\pi} \int_0^\pi \phi(a+2b\cos x) \,dx.
\end{equation}
\end{cor}

\noindent \textbf{Remark:} Recently, Trench \cite{MR2013470} obtained weaker conditions under which the conclusion of Corollary \ref{cor ks} holds. However, his conditions are of different nature as they are related to the spectrum of $J(a_k,b_k)$ rather than the sequences $(a_k)_k$ and $(b_k)_k$.

\subsection{The unbounded case: Part I}

We now turn our attention to the unbounded case, i.e. we now assume that the sequences $(\mathbf{a}^k)_k$ and $(\mathbf{b}^k)_k$ can take values over the whole real line. Our first result is concerned with sequences that can be contracted in order to fit the bounded case. The next definition is based on the terminology introduced by Van Assche \cite{MR903848}.

\begin{defn} \label{reg}
We say that $r:(0,\infty) \to (0,\infty)$ is a regularly varying function (at infinity) for the sequence $(\mathbf{a}^k,\mathbf{b}^k)_k$ if the normalized sequences
$$ \left(  \frac{\mathbf{a}^k}{r(k)} \right)_k, \ \left( \frac{\mathbf{b}^k}{r(k)} \right)_k$$
are in $\mathcal{S}$.
\end{defn}

\medskip

The next result is an immediate consequence of Theorem \ref{theo 1}, so its proof is omitted.

\begin{cor} \label{cor reg 1}
Let $r$ be a regularly varying function for the sequence $(\mathbf{a}^k,\mathbf{b}^k)_k$. Suppose that and $(\mathbf{a}^k/r(k), \, \mathbf{b}^k/r(k))_k$ is  $\mu$-distributed for some probability measure $\mu$ on $I^2$. Then,, we have
\[ \lim_{k \to \infty}\frac{1}{k} \text{Trace}\left[\phi\left(J\left( \frac{\mathbf{a}^k}{r(k)}, \,\frac{\mathbf{b}^k}{r(k)} \right) \right) \right] = \frac{1}{\pi} \int_{I^2} \int_0^\pi \phi(x+2y\cos t) \, dt \, d\mu(x,y)\]
for any $\phi \in C[-3,3]$
\end{cor}

\medskip

\noindent \textbf{Example 7:} Using the three-terms recurrence relation for the Hermite polynomials, it is well-known that the zeros $z_1^k,...,z_k^k$ of the $k$th degree Hermite polynomial are the eigenvalues of the tridiagonal matrix
\begin{equation}
H_k=\left(
\begin{array}{ccccc}
0 & 1/2 & 0 & \cdots & 0 \\
1 & 0 & 1/2 & \ddots & \vdots \\
0 & 2 & 0 & \ddots & 0 \\
\vdots & \ddots & \ddots & \ddots & 1/2 \\
0 & \cdots & 0 & k-1 & 0 \\
\end{array}
\right).
\end{equation}
It easy to see that $H_k$ is similar to the Jacobi matrix
\begin{equation}
H_k'=\left(
\begin{array}{ccccc}
0 & 1/\sqrt{2} & 0 & \cdots & 0 \\
1/\sqrt{2} & 0 & 1 & \ddots & \vdots \\
0 & 1 & 0 & \ddots & 0 \\
\vdots & \ddots & \ddots & \ddots & \sqrt{\frac{k-1}{2}} \\
0 & \cdots & 0 & \sqrt{\frac{k-1}{2}} & 0 \\
\end{array}
\right).
\end{equation}
Indeed, $H_k'=D_kH_kD_k^{-1}$ where $D_k=\text{diag}(d_1,...,d_{k})$ is the diagonal matrix whose elements are defined recursively by the relations
\[ d_1=1, \text{ and } d_{i+1}=\frac{d_i}{\sqrt{2i}} \qquad \text{ for } i=1,...,k-1. \]
The regularly varying function is $r(k)=\sqrt{2k}$ and the sequence
\[ \left( \frac{\mathbf{a}^k}{r(k)},\frac{\mathbf{b}^k}{r(k)} \right)_k = \left( \mathbf{0}^k, \left( \sqrt{\frac{j}{k}} \right)_{j=1}^k \right)_k \]
is $\mu$-distributed on $I^2$ with
\[ d\mu(x,y)= 2 y \delta_0(x) \, dx dy. \]
Consequently, the asymptotic distribution of the contracted zeros of Hermite polynomials is given by
\begin{eqnarray*}
\lim_{k \to \infty} \sum_{j=1}^k \phi \left( \frac{z_j^k}{\sqrt{2k}} \right) & = & \frac{1}{\pi} \int_{I^2} \int_0^\pi \phi(x+2y\cos t) dt \, d\mu(x,y) \\
  & = & \frac{2}{\pi} \int_0^1 \int_0^\pi \phi(2y\cos t) \, y \, dt \, dy\\
  & = & \frac{1}{2\pi} \int_{-2}^2 \phi(x) \sqrt{4-x^2} \, dx
\end{eqnarray*}
for any $\phi \in C[-\sqrt{2},\sqrt{2}]$. We thus obtain another proof of the well-known fact that the asymptotic distribution of the contracted zeros of Hermite polynomials is the Wigner semicircle distribution.

\medskip

\noindent \textbf{Example 8:} One can use a similar approach to derive the well-known fact that the asymptotic distribution of the contracted zeros of Laguerre polynomials is the Marchenko-Pastur distribution. Indeed, the zeros $z_1^k,...,z_k^k$ of the $k$th degree Laguerre polynomial are the eigenvalues of the Jacobi matrix
\begin{equation}
L_k=\left(
\begin{array}{ccccc}
1 & 1 & 0 & \cdots & 0 \\
1 & 3 & 2 & \ddots & \vdots \\
0 & 2 & 5 & \ddots & 0 \\
\vdots & \ddots & \ddots & \ddots & k-1 \\
0 & \cdots & 0 & k-1 & 2k-1 \\
\end{array}
\right).
\end{equation}
Hence, the regularly varying function is  $r(k)=k$ and the sequence is
\[ \left( \frac{\mathbf{a}^k}{r(k)},\frac{\mathbf{b}^k}{r(k)} \right)_k = \left( \left(\frac{2j-1}{k}\right)_{j=1}^{k}, \left( \frac{j}{k} \right)_{j=1}^{k} \right)_k \]
is $\mu$-distributed on $I^2$ with
\[ d\mu(x,y)= \delta_0(x-2y) \, dx dy. \]
The asymptotic distribution of the contracted zeros of Laguerre polynomials is therefore given by
\begin{eqnarray*}
\lim_{k \to \infty} \sum_{j=1}^k \phi \left( \frac{z_j^k}{k} \right) & = & \frac{1}{\pi} \int_{I^2} \int_0^{\pi} \phi(x+2y\cos t) \, dt \, d\mu(x,y) \\
 & = & \frac{1}{\pi} \int_0^1 \int_0^\pi \phi(2y+2y\cos t) \, dt \, dy\\
 & = & \frac{1}{2 \pi} \int_0^4 \phi(x) \ \frac{\sqrt{4x-x^2}}{x} \, dx
\end{eqnarray*}
for any $\phi \in C[0,4]$.

\medskip

More generally, we can use our trace formula in Corollary \ref{cor reg 1} to give a new proof of the following result due to Van Assche.

\begin{theo}
Let $r$ be a regularly varying function with parameter $\alpha \in (0,1)$, i.e. for every $t>0$,
\[ \lim_{x \to \infty} \frac{r(tx)}{r(x)} = t^{\alpha}. \]
Let $(a_k)_k$ and $(b_k)_k$ be two sequences with $a_k \in \R$ and $b_k>0$ satisfying the conditions
\begin{equation} \label{conditions}
  \lim_{k \to \infty} \frac{a_k}{r(k)} = a, \qquad \lim_{k \to \infty} \frac{b_k}{r(k)}=b.
\end{equation}
If $J(\mathbf{a}^k,\mathbf{b}^k)$ is the Jacobi matrix with $\mathbf{a}^k=(a_1,...,a_k)$ and $\mathbf{b}^k=(b_1,...,b_k)$, then we have
\begin{equation*}
  \lim_{k \to \infty} \frac{1}{k}  \text{Trace}\left[\phi\left(J\left( \frac{\mathbf{a}^k}{r(k)}, \,\frac{\mathbf{b}^k}{r(k)} \right) \right) \right]  = \frac{1}{\pi}\int_{a-2b}^{a+2b} f(x) \, v_\alpha(x) \, dx.
\end{equation*}
Here $v_\alpha(x)$ denotes the Nevai-Ullman density with parameter $\alpha$ defined as the Mellin convolution
\[ v_\alpha(x)=b_\alpha \ast \omega_{a,b}(x) = \int_0^1 b_\alpha(y) \omega_{a,b}(x/y) \,\frac{dy}{y} \]
with
\[ b_\alpha(y) = \begin{cases}
                    \alpha y^{\alpha-1} & \text{ if } y \in (0,1),\\
                    0 & \text{ elsewhere},
                 \end{cases} \]
and
\[ \omega_{a,b}(y) = \begin{cases}
       \frac{1}{\pi} \, \frac{1}{\sqrt{(a+2b-y)(y-a+2b)}} & \text{ if } y \in (a-2b,a+2b).\\
       0 & \text{elsewhere.}
       \end{cases} \]
\end{theo}

\medskip

\noindent \textit{Proof:} The conditions \eqref{conditions} imposed on the sequence $(a_k)_k$ and $(b_k)_k$ imply that for any $\epsilon>0$,
\[ \frac{ \# \left\{ 1\leq j \leq k: \left| \frac{a_j}{r(k)} - a \left( \frac{j}{k} \right)^{\alpha} \right| < \epsilon, \right\}}{k} \to 1 \]
and
\[ \frac{ \# \left\{ 1\leq j \leq k:  \, \left| \frac{b_j}{r(k)} - b \left( \frac{j}{k} \right)^{\alpha} \right| < \epsilon \right\}}{k} \to 1 \]
for $k \to \infty$. In particular, it follows that
\begin{equation*}
  \lim_{k \to \infty} \frac{1}{k}  \text{Trace}\left[\phi\left(J\left( \frac{\mathbf{a}^k}{r(k)}, \,\frac{\mathbf{b}^k}{r(k)} \right) \right) \right] =
  \lim_{k \to \infty} \frac{1}{k}  \text{Trace}\left[\phi\left(J\left( a\mathbf{u}^k, \, b\mathbf{u}^k \right) \right) \right]
\end{equation*}
where the sequence $\mathbf{u}^k$ is given by
\[ \mathbf{u}^k=(1/k^\alpha,(2/k)^\alpha,...,((k-1)/k)^\alpha,1). \]
Clearly, $\mathbf{u}^k \in \mathcal{S}$ and by Example 4, the sequence $\left( a\mathbf{u}^k, \, b\mathbf{u}^k \right)_k$ is $\mu$-distributed with $\mu$ satisfying
\[ \int_{I^2} \psi(x,y) \,d\mu(x,y) = \int_0^1 \psi(ax^\alpha,bx^\alpha) \, dx \]
for any $\psi \in C(I^2)$. By Theorem \ref{theo 1}, we deduce that
\begin{eqnarray*}
\lefteqn{\lim_{k \to \infty} \frac{1}{k}  \text{Trace}\left[\phi\left(J\left( \frac{\mathbf{a}^k}{r(k)}, \,\frac{\mathbf{b}^k}{r(k)} \right) \right) \right]}\\
 & & \qquad \qquad =  \frac{1}{\pi} \int_0^1 \int_0^{\pi} \phi(ax^\alpha+2bx^\alpha \cos (\pi t)) \, dt \,dx\\
 & & \qquad \qquad =  \frac{1}{\pi} \int_0^1 \int_{a-2b}^{a+2b} f(t^\alpha x) \frac{dx \, dt}{\sqrt{(a+2b-x)(x-a+2b)}}
\end{eqnarray*}
Under some simple changes of variable, last is easily seen to be equivalent to the expression given in the statement of theorem. $\qed$

\subsection{The unbounded case: Part II}

Our second result for Jacobi matrices for unbounded sequences is concerned with sequences $(\mathbf{a}^k)_k$ and $(\mathbf{b}^k)_k$ that belong to $\mathcal{S}'$ (see Definition \ref{unbounded}). Before we can go further, we need to reformulate Definition \ref{mu-distributed} for unbounded sequences.

\begin{defn} \label{mu-distributed unb}
We say that a sequence $(\mathbf{a}^k,\mathbf{b}^k)_k$ with $(\mathbf{a}^k)_k$ and $(\mathbf{b}^k)_k$ in $\mathcal{S}'$ is $\mu$-distributed if for any $\psi \in C_b(\R^2):=C(\R^2) \cap L_\infty(\R^2)$, the space of bounded continuous functions on $\R^2$, one has
\[ \lim_{k \to \infty} \frac{1}{k} \sum_{j=1}^k \psi(a^k_j,b^k_j) = \int_{\R^2} \psi(x,y) \ d\mu(x,y) \]
for some probability measure $\mu$ on $\R^2$.
\end{defn}

Recall, a probability measure is said to be tight if for any $\epsilon>0$, one can find a closed bounded rectangle $\mathcal{R}=[\alpha_1,\beta_1] \times [\alpha_2,\beta_2]$ such that
$$\mu(\R^2-\mathcal{R}) < \epsilon.$$

\medskip

\begin{theo} \label{theo unbounded}
Let $(\mathbf{a}^k)_k$ and $(\mathbf{b}^k)_k$ be two sequences in $\mathcal{S}'$. If $(\mathbf{a}^k,\mathbf{b}^k)_k$ is $\mu$-distributed for some tight probability measure $\mu$ on $\R^2$, then we have
\begin{equation} \label{unb 0}
\lim_{k \to \infty} \frac{1}{k}\text{Trace}\left[ \phi(J(\mathbf{a}^k,\mathbf{b}^k)) \right]
= \frac{1}{\pi} \int_{\R^2} \int_0^\pi \phi(x+2y \, \cos t)  \, d\mu(x,y)
\end{equation}
for any $\phi \in C_b(\R)$.
\end{theo}

\medskip

\noindent \textit{Proof:} By Proposition \ref{prop 2}, we know that
\begin{equation} \label{unb1}
\frac{1}{k}\text{Trace}\left[ J^n(\mathbf{a}^k,\mathbf{b}^k) \right]
= \frac{1}{k}\sum_{j=0}^{\lfloor n/2 \rfloor} \frac{n!}{(j!)^2(n-2j)!} \sum_{i=1}^k a_i^{n-2j} b_i^{2j} +  o(1).
\end{equation}
Since $\mu$ is tight and $(\mathbf{a}^k,\mathbf{b}^k)_k$ is $\mu$-distributed, there exists for any $\epsilon>0$ a closed bounded rectangle $\mathcal{R} \subseteq \R^2$ such that
\begin{equation} \label{unb1.1}
  \frac{\#\{ j \leq k: (a_j^k,b_j^k) \notin \mathcal{R} \}}{k} < \epsilon
\end{equation}
for $k$ large enough. We now introduce the smooth cutoff function $\theta:\R^2 \to [0,1]$ defined by
$$\theta(x,y) = \begin{cases}
           1  & \text{ if } (x,y) \in \mathcal{R}\\
           0 & \text{ if } (x,y) \notin \mathcal{R}_\epsilon
            \end{cases}$$
where $\mathcal{R}_\epsilon$ is a compact subset of $\R^2$ with $\mathcal{R} \subseteq \mathcal{R}_\epsilon$ and $m(\mathcal{R}_\epsilon/\mathcal{R})<\epsilon$. Equations \eqref{unb1} and \eqref{unb1.1} yield
\begin{multline} \label{unb2}
\frac{1}{k}\text{Trace}\left[ J^n(\mathbf{a}^k,\mathbf{b}^k) \right] \\
= \frac{1}{k}\sum_{j=0}^{\lfloor n/2 \rfloor} \frac{n!}{(j!)^2(n-2j)!} \sum_{i=1}^k (a_i^k)^{n-2j} (b_i^k)^{2j} \theta(a_i,b_i) +  o(1)
\end{multline}
for any nonnegative integer $n$. Moreover, $(\mathbf{a}^k,\mathbf{b}^k)_k$ being $\mu$-distributed implies
\begin{multline} \label{unb3}
\frac{1}{k}\text{Trace}\left[ J^n(\mathbf{a}^k,\mathbf{b}^k) \right] \\
= \sum_{j=0}^{\lfloor n/2 \rfloor} \frac{n!}{(j!)^2(n-2j)!} \int_{\R^2} x^{n-2j} y^{2j} \ \theta(x,y) \, d\mu(x,y)+ o(1).
\end{multline}
We now perform our usual trick, that is we introduce the $sinc$ function in \eqref{unb3} and use the multinomial theorem to obtain
\begin{multline} \label{unb4}
\frac{1}{k}\text{Trace}\left[ J^n(\mathbf{a}^k,\mathbf{b}^k) \right] \\
= \frac{1}{\pi} \int_{\R^2} \int_0^\pi (x+2y \, \cos t)^n \ \theta(x,y) \, d\mu(x,y)+ o(1).
\end{multline}
Note that 
$$ K_1:=\{ x+2y \cos t : (x,y) \in \mathcal{R}_\epsilon \text{ and } t \in [0,\pi] \}$$
is a compact subset of $\R$. Moreover, the assumptions on $\mu$ together with \eqref{unb1.1} imply that there exists a compact subset $K_2 \subseteq \R$ such that
$$ \# \{ \lambda : \lambda \in \sigma(J(\mathbf{a}^k,\mathbf{b}^k)), \ \lambda \notin K_2\} =o(k)  $$
as gets $k$ arbitrary large. In particular, we have
\begin{equation}
  \frac{1}{k}\text{Trace}\left[ J^n(\mathbf{a}^k,\mathbf{b}^k) \right] = \frac{1}{k} \sum_{\lambda \in \sigma(J(\mathbf{a}^k,\mathbf{b}^k)) \cap K_2} \lambda^n +o(1).
\end{equation}

By Weierstrass Approximation Theorem applied tp a closed bounded interval that contains $K_1 \cup K_2$, it is then easy to see that \eqref{unb4} yields
\begin{multline}
\frac{1}{k}\text{Trace}\left[ \phi(J(\mathbf{a}^k,\mathbf{b}^k)) \right] \\
= \frac{1}{\pi} \int_{\R^2} \int_0^\pi \phi(x+2y \, \cos t) \ \theta(x,y) \, d\mu(x,y)+ o(1)
\end{multline}
for any $\phi \in C_b(\R)$. Furthermore, since the measure $\mu$ is  tight, it follows that
\begin{eqnarray} \label{unb5}
\lefteqn{\left| \int_{\R^2}  \phi(x+2y \, \cos t) \ (1-\theta(x,y)) \, d\mu(x,y) \right|} \nonumber \\
& & \qquad \leq  \left| \int_{\R^2-R_\epsilon}  \phi(x+2y \, \cos t)  \, d\mu(x,y) \right| +  \nonumber\\
& & \qquad \quad \left| \int_{R_\epsilon-R}  \phi(x+2y \, \cos t) \ (1-\theta(x,y)) \, d\mu(x,y) \right| \nonumber\\
& & \qquad \leq 2 \|\phi\|_\infty \epsilon.
\end{eqnarray}
The desired trace formula \eqref{unb 0} is then an immediate consequence of \eqref{unb4} and \eqref{unb5}.  $\qed$

\medskip

\noindent \textbf{Example 9:} Let $(\mathbf{X}^k)_k$ and $(\mathbf{Y}^k)_k$ be two sequences with $\mathbf{X}^k$ and $\mathbf{Y}^k$ being $k$-vectors whose components are i.i.d. random variables that are normally distributed with mean zero and variance $\sigma^2(k) \to 0$ as $k \to \infty$. First, we note that for all $1 \leq j \leq k$,
\begin{multline} \label{erfc}
\text{Prob}\{ \omega \in \R: |X^k_j(\omega)|> \log k\} \\
= \frac{1}{\sqrt{2}} \, \text{Erfc} \, (\sigma(k) \, \log k )=  \Oo \left( \frac{e^{-\log^2 k/2\sigma^2(k)}}{k \log k} \right).
\end{multline}
Secondly, Chebyshev's inequality implies that
\begin{eqnarray*}
 \lefteqn{\text{Prob} \left\{ \omega \in \R:  |X_{j+1}^k(\omega)-X_j^k(\omega)|> \sigma^{1/2}(k) \right\}} \nonumber\\
  & & \leq  \frac{1}{\sigma(k)}  \int_{-\infty}^\infty \left| X_{j+1}^k(\omega)-X_j^k(\omega) \right|^2 \, e^{-\omega^2/2\sigma^2(k)} \, \frac{d\omega \nonumber}{\sqrt{2\pi}} \\
  & & \leq  \frac{8}{\sigma(k)}  \int_{-\infty}^\infty |X_j^k(\omega)|^2 \,  e^{-\omega^2/2\sigma^2(k)} \, \frac{d\omega \nonumber}{\sqrt{2\pi}}\\
  & & = 8 \sigma(k).
\end{eqnarray*}
From this, we easily deduce that
\begin{equation} \label{cheb}
\frac{1}{k} \sum_{j=1}^k |X_{j+1}^k(\omega)-X_j^k(\omega)| = \Oo(\sigma^{1/2}(k))
\end{equation}
for asymptotically almost every (a.a.e.) $\omega \in \R$. In particular, if we assume that $\sigma(k)=\Oo(k^{-\delta})$ for some $\delta>0$, it then follows from \eqref{erfc} and \eqref{cheb} that $(\mathbf{X}^k(\omega))_k \in \mathcal{S}'$ for a.a.e. $\omega \in \R$. Of course, a similar conclusion also hold for $(\mathbf{Y}^k(\omega))_k$.

Furthermore, under the independence assumption, one can apply the strong Law of Large Numbers to conclude
\begin{eqnarray*}
  \frac{1}{k} \sum_{j=1}^k  \psi(X_j(\omega),Y_j(\omega))
   &  = & \frac{1}{2\pi \sigma(k)} \int_{\R^2} \psi(x,y) \, e^{-(x^2+y^2)/2\sigma^2(k)} \, dx \, dy + o(1)\\
   &  = & \psi(0,0) +o(1)
\end{eqnarray*}
for a.a.e. $\omega \in \R$ and any $\psi \in C_b(\R^2)$. In other words, the sequence $(X_k(\omega),Y_k(\omega))_k$ is $\delta_0 \times \delta_0$-distributed in the sense of Definition \ref{mu-distributed unb}. By Theorem \ref{theo unbounded}, we conclude that for a.a.e. $\omega \in \R$,
\begin{equation*}
  \lim_{k \to \infty} \frac{1}{k} \text{Trace}[\phi(J(\mathbf{X}^k(\omega),\mathbf{Y}^k(\omega)))] = \phi(0)
\end{equation*}
for any $\phi \in C_b(\R)$.

\medskip

\subsection{The monotone case}

An obvious, but interesting corollary of Propositions \ref{prop 1} and \ref{prop 2} is obtained by considering sequences that are almost everywhere monotone.

\begin{defn}
We say that $(\mathbf{a}^k)_{k \in \N}$ is a.e. increasing if
\[ \frac{\#\{ 1 \leq i \leq k: a_i^k \leq a_{i+1}^k \}}{k} \to 1 \qquad  \text{as } k \to \infty. \]
\end{defn}

\medskip

Of course, a similar condition holds for a.e. decreasing sequences. Clearly, any a.e. monotone sequence that satisfies conditions (ii) or (ii') of Definitions \ref{def1} and \ref{unbounded} respectively will also satisfy conditions (i) or (i') of the same definitions. As a consequence, we obtain the following result.

\begin{prop}
Let $(\mathbf{a}^k)_k, (\mathbf{b}^k)_k$ be two a.e. monotone sequences with either
$$(\mathbf{a}^k)_k, (\mathbf{b}^k)_k \in \mathcal{S}$$
or
$$(\mathbf{a}^k)_k, (\mathbf{b}^k)_k \in \mathcal{S}'.$$
Then, for any $n \in \N$, we have
\begin{equation} \label{trace0}
\text{Trace}\left[ J^n(\mathbf{a}^k,\mathbf{b}^k) \right]=\sum_{j=0}^{\lfloor n/2 \rfloor} \frac{n!}{(j!)^2(n-2j)!} \sum_{i=1}^k (a_i^k)^{n-2j} (b_i^k)^{2j} +  o(k).
\end{equation}
\end{prop}

If we assume furthermore that the sequence $(\mathbf{a}^k,\mathbf{b}^k)_k$ is $\mu$-distributed as in the sense of Definitions \ref{mu-distributed} or \ref{mu-distributed unb}, then we easily derive the following two results that are simple consequence of Theorems \ref{theo 1} and \ref{theo unbounded}.

\begin{cor} \label{cor mono 1}
Let $(\mathbf{a}^k)_k$ and $(\mathbf{b}^k)_k$ be two a.e. monotone sequences in $[0,1]$. If $(\mathbf{a}^k,\mathbf{b}^k)_k$ is $\mu$-distributed for some probability measure $\mu$ on $I^2$, then we have
\begin{equation}
\lim_{k \to \infty} \frac{1}{k}\text{Trace}\left[ \phi(J(\mathbf{a}^k,\mathbf{b}^k)) \right]
= \frac{1}{\pi} \int_{I^2} \int_0^\pi \phi(x+2y \, \cos t)  \, d\mu(x,y)
\end{equation}
for any $\phi \in C[-3,3]$.
\end{cor}

\begin{cor} \label{cor mono 2}
Let $(\mathbf{a}^k)_k$ and $(\mathbf{b}^k)_k$ be two a.e. monotone sequences in $\R$ that satisfy the estimates
$$ |a_j^k|=\Oo(\log k) \text{ and } |b_j^k|=\Oo(\log k) \text{ for all } 1 \leq j \leq k, \ k \in \N.$$
If $(\mathbf{a}^k,\mathbf{b}^k)_k$ is $\mu$-distributed for some tight probability measure $\mu$ on $\R^2$, then we have
\begin{equation}
\lim_{k \to \infty} \frac{1}{k}\text{Trace}\left[ \phi(J(\mathbf{a}^k,\mathbf{b}^k)) \right]
= \frac{1}{\pi} \int_{\R^2} \int_0^\pi \phi(x+2y \, \cos t)  \, d\mu(x,y)
\end{equation}
for any $\phi \in C_b(\R)$.
\end{cor}

\medskip

\noindent \textbf{Example 10:} Let $(\mathbf{X}^k)_k$ and $(\mathbf{Y}^k)_k$ be sequences where
$$\mathbf{X}^k=(X_1^k,...,X_k^k) \text{ and } \mathbf{Y}^k=(Y_1^k,...,Y_k^k)$$
are two $k$ random vectors. We assume for all $1\leq j \leq k$, $k \in \N$, that the $X_j^k$'s and $Y_j^k$'s are i.i.d. random variables of mean zero and variance one on some probability space $(\Omega,P)$. We denote by $\mathbf{X}^{(k)}$ the random vector whose $j$th entry is the $j$th order statistic $X_{(j)}^k$ obtained from $X_1^k,...,X_k^k$, i.e.
$$ \mathbf{X}^{(k)}=(X_{(1)}^k,...,X_{(k)}^k)$$
and similarly for $\mathbf{Y}^{(k)}$. For any $\omega \in \Omega$, we denote by $J_k(\omega)$ the Jacobi matrix given by
\begin{equation}
\mathbf{J}_{(k)}(\omega) = \left(
\begin{array}{ccccc}
X_{(1)}^k(\omega) & Y_{(1)}^k(\omega) & 0 & \cdots & 0 \\
Y_{(1)}^k(\omega) & X_{(2)}^k(\omega) & Y_{(2)}^k(\omega) & \ddots & \vdots \\
0 & Y_{(2)}^k(\omega) & X_{(3)}^k(\omega) & \ddots & 0 \\
\vdots & \ddots & \ddots & \ddots & Y_{(k-1)}^k(\omega) \\
0 & \cdots & 0 & Y_{(k-1)}^k(\omega) & X_{(k)}^k(\omega) \\
\end{array}
\right).
\end{equation}
By definition of order statistics, the sequences $\mathbf{X}^{(k)}$ and $\mathbf{Y}^{(k)}$ are monotone increasing. If we assume that $X_j^k(\omega) \in [0,1]$ and $Y_j^k(\omega) \in (0,1]$, then the strong Law of Large Numbers implies that for a.a.e $\omega \in \Omega$,
\begin{eqnarray*}
  \lim_{k \to \infty} \sum_{j=1}^k \psi(X_{(j)}^k(\omega),Y_{(j)}^k(\omega)) = \int_{I^2} \psi(x,y) \ d\mu_{X,Y}(x,y)
\end{eqnarray*}
for any $\psi \in C(I^2)$. Here, we denote by $\mu_{X,Y}$ the joint probability distribution of $X$ and $Y$ where $X$ and $Y$ are two random variables on $\Omega$ that follow the same distribution as $X_j^k$ and $Y_j^k$ respectively. Consequently, we can apply Corollary \ref{cor mono 1} to conclude that
\begin{equation}
\lim_{k \to \infty} \frac{1}{k}\text{Trace}\left[ \phi(\textbf{J}_k(\omega) \right]
= \frac{1}{\pi} \int_{I^2} \int_0^\pi \phi(x+2y \, \cos t)  \, d\mu_{X,Y}(x,y)
\end{equation}
for any $\phi \in C[-3,3]$ and a.a.e $\omega \in \Omega$.

Similarly, if we assume that $X_j^k = \Oo(\log k)$ and $Y_j^k= \Oo(\log k)$, then we can apply Corollary \ref{cor mono 2} to obtain
\begin{equation}
\lim_{k \to \infty} \frac{1}{k}\text{Trace}\left[ \phi(\textbf{J}_k(\omega) \right]
= \frac{1}{\pi} \int_{\R^2} \int_0^\pi \phi(x+2y \, \cos t)  \, d\mu_{X,Y}(x,y)
\end{equation}
for any $\phi \in C_b(\R)$ and a.a.e $\omega \in \Omega$.

\section{Existence Results}

In this section, we would like to address the following question: Given two sequences $(\mathbf{a}^k)_k$ and $(\mathbf{b}^k)_k$ in $\mathcal{S}$ or $\mathcal{S}'$, does there exist a probability measure $\mu$ with support in $I^2$ or a tight measure probability measure $\mu$ with support in $\R^2$ for which the sequence $(\mathbf{a}^k,\mathbf{b}^k)_k$ is $\mu$-distributed? It is easy to construct examples of sequences that are not $\mu$-distributed for any probability measure. For instance, take the sequence $(\mathbf{a}^k,\mathbf{b}^k)_k$ with
\[ (a_j^k,b_j^k) = \begin{cases}
(0,1) & \text{ if } 2 | k \\
(1,1) & \text{ if } 2 \nmid k.
\end{cases} \]
However, as the next result shows one can always find subsequences that are $\mu$-distributed.

\begin{prop} \label{prop ex 1}
Let $(\mathbf{a}^k)_k$ and $(\mathbf{b}^k)_k$ be two sequences in $\mathcal{S}$. Then, there exists a subsequence of positive integers $k_j$ and a probability measure $\mu$ with support in $[0,1]^2$ such that $(\mathbf{a}^{k_j},\mathbf{b}^{k_j})_j$ is $\mu$-distributed.
\end{prop}

\noindent \textit{Proof:} For any $x,y \in \R$, consider the distribution functions $F_k$  defined by
$$F_k(x,y) = \frac{\# \{ j \leq k: a_j^k < x \text{ and } b_j^k<y\}}{k}.$$
By Helly's selection principle (\cite{MR830424}, Theorem 25.9), we can find a subsequence of positive integers $k_j$, $k_j \to \infty$ as $j \to \infty$, such that $F_{k_j}$ converges weakly (in the probability sense) to a distribution function $F$. By Theorem 12.5 in \cite{MR830424}, we can associate to $F_{k_j}$ and $F$ unique probability measures $\mu_{k_j}$ and $\mu$ on $[0,1]^2$  defined by
\[ \mu((a,b) \times (c,d)) =F(b,d)-F(b,c)-F(a,d)+F(a,c) \]
and similarly for $\mu_{k_j}$. Since $\mu_{k_j}$ converges weakly to $\mu$, we can apply Theorem 29.1 in \cite{MR830424} to conclude that $(\mathbf{a}^{k_j},\mathbf{b}^{k_j})$ is $\mu$-distributed as in the sense of Definition \ref{mu-distributed}. $\qed$

\medskip

\begin{prop} \label{prop ex 2}
Let $(\mathbf{a}^k)_k$ and $(\mathbf{b}^k)_k$ be two sequences in $\mathcal{S}'$. Suppose that for any $\epsilon>0$, there exists a bounded rectangle $R$ such that
\begin{equation} \label{tight seq}
 \frac{\# \{ 1 \leq j \leq k : |a_j^k| \notin R\}}{k} < \epsilon, \  \text{ and } \  \frac{\# \{ 1 \leq j \leq k : |b_j^k| \notin R\}}{k} < \epsilon 
\end{equation}
for $k$ large enough. Then, there exists a subsequence of positive integers $k_j$ and a tight probability measure $\mu$ on $\R^2$ such that $(\mathbf{a}^{k_j},\mathbf{b}^{k_j})_j$ is $\mu$-distributed.
\end{prop}

\noindent \textit{Proof:} The argument is identical as for the proof of Proposition \ref{prop ex 1} with the exception that Theorem 29.1 in \cite{MR830424} must be replaced by Theorem 29.3. $\qed$

\medskip

By combining Proposition \ref{prop ex 1} and Theorem \ref{theo 1}, or Proposition \ref{prop ex 2} together with Theorem \ref{theo unbounded}, we obtain the following two existence results.

\begin{cor}  \label{subsequence}
Let $(\mathbf{a}^k)_k$ and $(\mathbf{b}^k)_k$ be two sequences in $\mathcal{S}$. Then, there exists a subsequence of positive integers $k_j$ and a probability measure $\mu$ with support in $I^2$ such that
\begin{equation} \label{trace 1 sub}
\lim_{j \to \infty} \frac{1}{k_j} \text{Trace}[\phi(J(\mathbf{a}^{k_j},\mathbf{b}^{k_j}))] = \frac{1}{\pi}\int_{I^2} \int_0^\pi \phi(x+2y\cos t) \, dt \, d\mu(x,y)
\end{equation}
for any $\phi \in C[-3,3]$.
\end{cor}

\begin{cor}
Let $(\mathbf{a}^k)_k, \ (\mathbf{b}^k)_k \in \mathcal{S}'$ be two sequences satisfying \eqref{tight seq}. Then, there exists a subsequence of positive integers $k_j$ and a probability measure $\mu$ such that
\begin{equation} \label{trace 1 sub unb}
\lim_{j \to \infty} \frac{1}{k_j} \text{Trace}[\phi(J(\mathbf{a}^{k_j},\mathbf{b}^{k_j}))] = \frac{1}{\pi}\int_{\R^2} \int_0^\pi \phi(x+2y\cos t) \, dt \, d\mu(x,y)
\end{equation}
for any $\phi \in C_b(\R)$.
\end{cor}

\section{Conclusion}

Our results establish new and very general conditions for which one can compute explicitly the asymptotic distribution of the spectrum of Jacobi matrices. Our results are new in the sense that we do not assume any type of convergence for the sequences $(\mathbf{a}^k)_k,(\mathbf{b}^k)_k$ of $J(\mathbf{a}^k,\mathbf{b}^k)$, but we only assume they satisfy a small-deviation condition and they are $\mu$-distributed.

A quick a look at the proofs should convince the readers that similar techniques as the ones developed in this paper can be used to compute an asymptotic trace formula for band matrices. For instance, it is straightforward to extend the proof of Propositions \ref{prop 1} and \ref{prop 2} to matrices with three diagonal bands of the form
\begin{equation}
 \left(
\begin{array}{cccccccc}
a_1^k & &   & &  & b_1^k & & \\
      & \ddots & &   & &  & \ddots &   \\
      &       & \ddots &  &  &  & & b_{k-q}^k\\
c_1^k &       &        & a_p^k &  &   & & \\
      & \ddots&        &       & \ddots    & &  & \\
      &       & \ddots &       &           & a_q^k & & \\
      &       &        & \ddots &           &      & \ddots & \\
      &       &        &       & c_{k-p}^k          &      &        & a_k^k
\end{array}
\right).
\end{equation}
where the sequences $(\mathbf{a}^k)_k,(\mathbf{b}^k)_k,(\mathbf{c}^k)_k$ are assumed to satisfy the usual small-deviation conditions. In particular, we would like to use these extensions to derive new results for the asymptotic distribution of difference and Sturm-Liouville operators. We hope to address these points in future research work.

\bibliographystyle{plain}

\begin{thebibliography}{10}

\bibitem{agbo07}
A.~Agnew and A.~Bourget.
\newblock Semi-classical density of states for the quantum asymmetric top.
\newblock {\em J. Phys. A: Math. Theor.}, 41(18), 2008.

\bibitem{agbo09}
Alfonso~F. Agnew and Alain Bourget.
\newblock A trace formula for a family of {J}acobi operators.
\newblock {\em Anal. Appl. (Singap.)}, 7(2):115--130, 2009.

\bibitem{MR830424}
Patrick Billingsley.
\newblock {\em Probability and measure}.
\newblock Wiley Series in Probability and Mathematical Statistics: Probability
  and Mathematical Statistics. John Wiley \& Sons Inc., New York, second
  edition, 1986.

\bibitem{bosh08}
J.~Borcea and B.~Shapiro.
\newblock Root asymptotics of spectral polynomials for the {L}am\'e operator.
\newblock {\em arXiv:math/0701883v1 (math-ca)}, 2007.

\bibitem{MR883643}
H.~L. Cycon, R.~G. Froese, W.~Kirsch, and B.~Simon.
\newblock {\em Schr\"odinger operators with application to quantum mechanics
  and global geometry}.
\newblock Texts and Monographs in Physics. Springer-Verlag, Berlin, study
  edition, 1987.

\bibitem{deift96}
P.~A. Deift.
\newblock {\em Orthogonal polynomials and random matrices: a
  {R}iemann-{H}ilbert approach}, volume~3 of {\em Courant Lecture Notes in
  Mathematics}.
\newblock New York University Courant Institute of Mathematical Sciences, New
  York, 1999.

\bibitem{MR956176}
Jeffrey~S. Geronimo, Evans~M. Harrell, II, and Walter Van~Assche.
\newblock On the asymptotic distribution of eigenvalues of banded matrices.
\newblock {\em Constr. Approx.}, 4(4):403--417, 1988.

\bibitem{MR1866252}
A.~B.~J. Kuijlaars and S.~Serra~Capizzano.
\newblock Asymptotic zero distribution of orthogonal polynomials with
  discontinuously varying recurrence coefficients.
\newblock {\em J. Approx. Theory}, 113(1):142--155, 2001.

\bibitem{kuva99}
A.B.J. Kuijlaars and W.~Van Assche.
\newblock The asymptotic zero distribution of orthogonal polynomials with
  varying recurrence coefficients.
\newblock {\em J. Approx. Theory}, 99:167--197, 1999.

\bibitem{MR0419394}
L.~Kuipers and H.~Niederreiter.
\newblock {\em Uniform distribution of sequences}.
\newblock Wiley-Interscience [John Wiley \& Sons], New York, 1974.
\newblock Pure and Applied Mathematics.

\bibitem{ne79}
P.G. Nevai.
\newblock {\em Orthogonal polynomials}, volume~18 of {\em Memoirs of the
  {A}merican Mathematical Society}.
\newblock American mathematical society, Providence, RI, 1979.

\bibitem{MR2013470}
William~F. Trench.
\newblock A note on asymptotic zero distribution of orthogonal polynomials.
\newblock {\em Linear Algebra Appl.}, 375:275--281, 2003.

\bibitem{MR903848}
Walter Van~Assche.
\newblock {\em Asymptotics for orthogonal polynomials}, volume 1265 of {\em
  Lecture Notes in Mathematics}.
\newblock Springer-Verlag, Berlin, 1987.

\end{thebibliography}

\end{document}